\documentclass[%
 superscriptaddress,
preprintnumbers,
 nofootinbib,
 amsmath,amssymb,
 aps,
twocolumn,
prl,
]{revtex4-2}

\usepackage{graphicx} % Required for inserting images
\usepackage{physics}
\usepackage{color}
\usepackage{hyperref}
\usepackage{comment}
\usepackage{array}
\usepackage{mathtools}

\hypersetup{
  colorlinks=true,
  linkcolor=[rgb]{0.60,0.00,0.00},
  citecolor=[rgb]{0.00,0.00,0.60},
  urlcolor=[rgb]{0.00,0.00,0.60},
  setpagesize=false
}

\newcommand{\SU}{\mathrm{SU}}

\begin{document}
\begin{titlepage}
\preprint{RIKEN-iTHEMS-Report-26, YITP-26-01}

\title{Phases of the $q$-deformed $\SU(N)$ Yang-Mills theory at large $N$}

\author{Tomoya Hayata}
\email{hayata@keio.jp}
\affiliation{Department of Physics, Keio University, 4-1-1 Hiyoshi, Yokohama, Kanagawa 223-8521, Japan}
\affiliation{RIKEN Center for Interdisciplinary Theoretical and Mathematical Sciences
(iTHEMS), RIKEN, Wako 351-0198, Japan}
\affiliation{International Center for Elementary Particle Physics and The University of Tokyo, 7-3-1 Hongo, Bunkyo-ku, Tokyo 113-0033, Japan}

\author{Yoshimasa Hidaka}
\email{yoshimasa.hidaka@yukawa.kyoto-u.ac.jp}
\affiliation{Yukawa Institute for Theoretical Physics, Kyoto University, Kitashirakawa Oiwakecho, Sakyo-ku, Kyoto 606-8502, Japan}
\affiliation{RIKEN Center for Interdisciplinary Theoretical and Mathematical Sciences
(iTHEMS), RIKEN, Wako 351-0198, Japan}

\author{Hiromasa Watanabe}
\email{hiromasa.watanabe@keio.jp}
\affiliation{Department of Physics, Keio University, 4-1-1 Hiyoshi, Yokohama, Kanagawa 223-8521, Japan}
\affiliation{Research and Education Center for Natural Sciences, Keio University, 4-1-1 Hiyoshi, Yokohama, Kanagawa 223-8521, Japan}

\date{\today}% It is always \today, today,

\begin{abstract}
We investigate the $(2+1)$-dimensional $q$-deformed $\SU(N)_k$ Yang-Mills theory in the lattice Hamiltonian formalism, which is characterized by three parameters: the number of colors $N$, the coupling constant $g$, and the level $k$.
By treating these as tunable parameters, we explore how key properties of the theory, such as confinement and topological order, emerge in different regimes.
Employing a variational mean-field analysis that interpolates between the strong- and weak-coupling regimes, we determine the large-$N$ phase structure in terms of the 't~Hooft coupling $\lambda_\mathrm{tH}=g^2N$ and the ratio $k/N$.
We find that the topologically ordered phase remains robust at large $N$ under appropriate scalings of these parameters.
This result indicates that the continuum limit of large-$N$ gauge theory may be more intricate than naively expected, and motivates studies beyond the mean-field theory,  both to achieve a further understanding of confinement in gauge theories and to guide quantum simulations of large-$N$ gauge theories.

\end{abstract}

\maketitle
\end{titlepage}

\section{Introduction}

Gauge theories with finite-dimensional Hilbert spaces have long been studied as effective quantum field theory descriptions of topological phases of quantum matter~\cite{Dijkgraaf:1989pz}.
More recently, they have attracted growing interest in the context of quantum simulations of gauge theories~\cite{Banuls:2019bmf,Wiese:2021djl,DiMeglio:2023nsa,Fontana:2024rux,Halimeh:2025vvp}, as well as in modern approaches to confinement and the vacuum structure of Quantum Chromodynamics (QCD) and QCD-like theories based on generalized symmetries~\cite{Gaiotto:2014kfa,Kapustin:2014gua}.
A prominent example, also addressed in this Letter, is the $q$-deformed $\SU(N)_k$ Yang–Mills (YM) theory in $(2+1)$ dimensions. 
This theory interpolates between a confining nonabelian gauge theory and a topological quantum field theory described by a unitary modular tensor category (UMTC)~\cite{Kitaev:2005hzj,Barkeshli:2014cna}.
The level $k$ truncates the spectrum of irreducible representations, introducing a spurious phase transition 
(in the context of conventional high-energy physics)
between confinement and topological order.

Truncating the Hilbert space of gauge fields by introducing a cutoff $k$ on irreducible representations enlarges the parameter space.
Specifically, the $q$-deformed $\SU(N)_k$ YM theory has three parameters: the number of colors $N$, the coupling constant $g$, and the level $k$.
Treating these as tunable parameters, one can explore how confinement or topological order emerges in different regimes.
Previous works, however, have focused on the cases $N=2$ and $3$~\cite{Biedenharn:1989jw,Cunningham:2020uco,Bonzom:2022bpv,Zache:2023dko,Hayata:2023puo,Hayata:2023bgh,Hayata:2024fnh}. 
For fixed $N$, the continuum limit has been discussed as a sequential limit, with $k\to\infty$ taken prior to $1/g^2\to\infty$.
Then, a central question addressed in this Letter is whether a meaningful large-$N$ limit of the $q$-deformed theory can be defined. 
Since the deformation introduces a natural tunable ratio $k/N$, this allows one to consider alternative large-$N$ limits other than the 't~Hooft limit~\cite{tHooft:1973alw}.
Taking $N \to \infty$ with fixed ’t Hooft coupling $\lambda_\mathrm{tH} = g^2 N$ and $k/N$ yields a rich phase structure that is inaccessible in the conventional continuum YM theory.

In this Letter, we analyze the ground state of the $q$-deformed $\SU(N)_k$ YM theory in the Hamiltonian formalism.
Employing a variational mean-field ansatz that interpolates between the strong-coupling confined vacuum and the string-net condensed state~\cite{Levin:2004mi} realized in the weak-coupling regime, we derive a stability criterion for the topological phase expressed in terms of fusion rules, quantum dimensions, and quadratic Casimirs:
Instabilities in the Hessian of the mean-field energy determine the critical coupling $g_{\mathrm{c}}^2(N,k)$ separating the topological and confined phases.
When expressed in terms of $(1/\lambda_\mathrm{tH}, k/N)$, the critical lines for $N \ge 3$ collapse onto a universal curve, indicating that the large-$N$ limit at fixed $\lambda_\mathrm{tH}$ and $k/N$ is well-defined at the mean-field level. 
These results provide quantitative evidence that the topologically ordered phase survives in the large-$N$ limit when $g$ and $k$ are scaled appropriately. 
This offers guidance for the design of large-$N$ gauge-theory quantum simulators; for related works, see also Refs.~\cite{Banerjee:2012xg,Ciavarella:2024fzw,Ciavarella:2025bsg,Buser:2020cvn,Bergner:2024qjl,Halimeh:2024bth,Hanada:2025goy}.

\section{$q$-deformation}

The $q$ deformation restricts the number of possible irreducible representations.
The irreducible representations $\lambda$ are labeled by the Young diagrams and run among the set 
$
    P_k \coloneqq \qty{
        \lambda \;\left|\; 
        k\ge\lambda_1 \ge \lambda_2 \ge \cdots \ge \lambda_{N-1} \ge \lambda_N = 0
        \right.
    }.
$
In other words, Young diagrams whose number of columns is less than or equal to $k$ are allowed.
The fact that the number of irreducible representations is $\abs{P_k} = \binom{k+N-1}{N-1}$  enables us to regulate the Hilbert space by the integer $k$ (see also Ref.~\cite{Bonatsos:1999xj}).

For irreducible representations, $\lambda$ and $\mu$, their fusion is defined in terms of the \emph{fusion coefficients} $N_{\lambda\mu}^{(k)\nu}$ as
\begin{equation}
    \lambda \times \mu = \sum_{\nu \in P_k} N_{\lambda\mu}^{(k)\nu}\, \nu.
\end{equation}
We will omit the superscript $(k)$ of the fusion coefficients from now on.
To calculate the fusion coefficients, we employ the Verlinde formula~\cite{Verlinde:1988sn} 
\begin{equation}
    N_{\lambda\mu}^\nu
    =
    \sum_{\sigma\in P_k} \frac{S_{\lambda \sigma}S_{\mu \sigma}S_{\nu \sigma}^{-1}}{S_{\emptyset\sigma}},
\end{equation}
with $\emptyset$ being the irreducible representation such that the total number of boxes in the Young diagram is zero, i.e.,  $\lambda_1=\cdots=\lambda_{N-1}=0$, and the modular $S$-matrix $S$~\cite{Kac:1984mq} (see the Supplemental Material for the details).
The fusion coefficients are nonnegative integers and satisfy $N_{\lambda\mu}^\nu = N_{\mu\lambda}^\nu = N_{\mu\bar\nu}^{\bar\lambda} = N_{\bar\lambda\bar\mu}^{\bar\nu}$.
In the limit $k\to\infty$, the fusion rules reduce to those of $\mathfrak{su}(N)$.
Recall that the Verlinde formula provides a powerful method for computing fusion coefficients in rational conformal field theories, such as the Wess–Zumino–Novikov–Witten models. 
It is quite intriguing that this class of theories, developed in a very different context, finds an unexpected point of contact with YM theory and QCD. 

The quantum dimension $d_\lambda$ is also an important algebraic quantity,
which can be computed by the modular $S$-matrix as $d_\lambda = S_{\lambda\emptyset}/S_{\emptyset\emptyset}$.
The quantum dimensions are not integers, in general, and are equal to those of their anti-representation, $d_\lambda = d_{\bar{\lambda}}$.
At $k\to\infty$, the quantum dimension also reduces to the Lie-algebraic dimension of $\lambda$.
We also utilize the total quantum dimension defined as
$D\coloneqq \sqrt{\sum_{\lambda\in P_k} d_\lambda^2}$.

In this work, we numerically evaluate the fusion coefficients for various $N$ and $k$. 
We mainly investigate points of $(N,k)$ such that $\abs{P_k} \lesssim 3000$ due to the computational cost of fusion coefficients.
The analysis code used in this work is available at Ref.~\cite{code_HW}.

\section{Mean-field analysis of $q$-deformed YM}

We first review how to treat the physical states in the Hamiltonian formalism~\cite{Robson:1981ws,Zache:2023dko,Hayata:2023bgh,Hayata:2023puo}.
The gauge-invariant physical Hilbert space of gauge theory on the lattice can be spanned by a basis corresponding to a network of Wilson lines.
Our primary interest is the $\SU(N)$ gauge theory in $(2+1)$ dimensions.
Let us consider its lattice regularized and $q$-deformed theory. 
The Hamiltonian of $\SU(N)_k$ theory is given through the Kogut-Susskind formalism~\cite{Kogut:1974ag} as 
\begin{equation}
    H_\mathrm{KS} 
    = 
    \frac{1}{2}\sum_{e\in\mathcal{E}} E^2(e) 
    -
    \frac{K}{2}\sum_{f\in\mathcal{F}} 
    \left(
    \operatorname{tr}U_\mathrm{fund}(f)
    +
    \operatorname{tr}U_{\overline{\mathrm{fund}}}(f)
    \right),
    \label{eq:H_KS}
\end{equation}
with a coupling $K = 1/g^4$ in lattice unit.
Here, $\mathcal{E}$ and $\mathcal{F}$ are the sets of edges and faces of the system, respectively.
The electric field operator $E(e)$ lives on edges $e$ of the square lattice $\mathcal{E}$, and $\tr U_\mathrm{fund}(f)$ is the Wilson loop operator in the fundamental representation that circles the edges of the plaquette $f$ clockwise.
The action of operators in Eq.~\eqref{eq:H_KS} can be represented in a graphical notation (for details, see e.g., Refs.~\cite{Hayata:2023puo,Hayata:2023bgh}).
Notice that the mean-field analysis does not require the explicit form of the $F$-symbols, whereas it becomes necessary when the Wilson loop operator is applied to a general state.

To proceed, we employ the variational wave function~\cite{Bonatsos:1999xj,Zache:2023dko}:
$
    \ket{\Psi} = \prod_f \sum_{\lambda_f}\psi(\lambda_f)\,\operatorname{tr}U_{\lambda_f}\ket{\Omega},
$
where 
$\ket{\Omega}$
is the vacuum state in the strong-coupling limit, satisfying $E^2(e)\ket{\Omega}=0$ for all $e\in\mathcal{E}$.
The variational parameter $\psi(\lambda_f)$ obeys the normalization condition $\mathcal{N}(\psi)=\sum_{\lambda_f} \abs{\psi(\lambda_f)}^2=1$, ensuring $\braket{\Psi}=1$.
It is advantageous that this ansatz covers the ground state of both the strong and weak coupling limits by tuning the variational parameters.
By assuming the translational symmetry under the open boundary conditions in the infinite volume limit, we take the same wave function on all plaquettes.
Now, the variational problem can be solved analytically, which leads to the mean-field Hamiltonian density~\cite{Hayata:2023bgh}
\begin{equation}
\begin{split}
    E[\psi,\psi^*]  
    &\coloneqq
    \frac{\mel{\Psi}{H_\mathrm{KS}}{\Psi}}{V}
    \\
    &= 
    \frac{1}{\mathcal{N}^2}
    \sum_{\lambda,\mu,\nu} C_2(\nu) N_{\lambda\nu}^\mu \frac{d_\nu}{d_\lambda d_\mu} \left|\psi(\lambda)\right|^2\left|\psi(\mu)\right|^2
    \\
    &\quad 
    -
    \frac{K}{2\mathcal{N}}\sum_{\lambda,\mu}\psi^*(\lambda)M_{\lambda\mu}\psi(\mu),
    \label{eq:Hamiltonian_MF}
\end{split}
\end{equation}
where $V=|\mathcal{F}|$ is the volume of the system (the number of plaquettes), $C_2(\lambda)$ denotes the quadratic Casimir invariant, and
$
M_{\lambda\mu} \coloneqq N_{\mathrm{fund}\,\mu}^\lambda + N_{\overline{\mathrm{fund}}\,\mu}^\lambda.
$
The fundamental representation, $\mathrm{fund}$, is specified by $\lambda_1=1$, $\lambda_2=\cdots=\lambda_{N-1}=0$.
The wave function $\psi(\lambda)$ that minimizes Eq.~\eqref{eq:Hamiltonian_MF} varies depending on the YM coupling $1/g^2$ and cutoff $k$.
Correspondingly, the phase structure is divided into two regions, the confined phase and the topological phase. 
The former is connected smoothly to the ground state in the strong-coupling limit $1/g^2 = 0$, whereas the latter is connected to the ground state in the weak-coupling limit $1/g^2\to\infty$ (at finite $k$).
The mean-field computation implies that the ground state in the topological phase is identical to the string-net condensed state~\cite{Levin:2004mi,Zache:2023dko}, i.e., $\psi_0(\lambda) = d_\lambda/D$, which gives the topological order characterized by the UMTC.
In order to take the limit achieving the $\SU(N)$ YM theory in the continuous spacetime, one should control the parameters $1/g^2\to\infty$ and $k\to\infty$ in a nontrivial manner, avoiding entering the topological phase.
To this end, it is quite important to understand the phase boundary.

Our strategy to determine the phase boundary for each $N$ is to investigate the stability of the mean-field theory around the topological phase by assuming the second-order phase transition.
We consider 
\begin{equation}
    \psi(\lambda) = \psi_0(\lambda) + \delta\psi(\lambda),
    \label{eq:psi_around_SNC}
\end{equation}
where $\psi_0(\lambda)$ is the wave function of the string-net condensed state, which is real.
Notice that $\mathcal{N}(\psi) > 1$ in the presence of $\delta\psi$, 
and we need to take the normalization into account. 
For complex $\delta\psi$, one finds that the second-order perturbation has a form 
\begin{equation}
    E^{(2)} = 
    \sum_{\lambda,\mu} \mathbf{v}(\lambda)^\dagger \mathcal{M}_{\lambda\mu}  \mathbf{v}(\mu),
\end{equation}
where $\mathbf{v}(\lambda)^\top = (\delta\psi(\lambda),\delta\psi^*(\lambda))$ and
\begin{gather}
\begin{split}
    \mathcal{M}_{\lambda\mu} 
    =
    \begin{pmatrix}
        H_{\psi^*(\lambda)\psi(\mu)}
        &
        H_{\psi^*(\lambda)\psi^*(\mu)}
        \\
        H_{\psi(\lambda)\psi(\mu)}
        &
        H_{\psi(\lambda)\psi^*(\mu)}
    \end{pmatrix},
    \label{eq:Hessian}   
    \\
    H_{\psi_1\psi_2}
    =
    \left.\frac{\partial^2 E}{\partial \psi_1 \partial\psi_2}\right|_{\psi=\psi_0}.
\end{split}
\end{gather}
After a straightforward calculation,
each component can be expressed as 
\begin{gather}
\begin{split}
    H_{\psi(\lambda)\psi(\mu)}
    =
    2\sum_{\nu} C_2(\nu) \frac{d_\nu}{D^2} 
    \left(N_{\lambda\nu}^\mu - \frac{d_\lambda d_\mu d_\nu}{D^2}\right),
    \\
    H_{\psi(\lambda)\psi^*(\mu)} 
    = 
    H_{\psi(\lambda)\psi(\mu)}
    -\frac{K}{2}\left(
    M_{\lambda\mu} - 2d_\mathrm{fund}\delta_{\lambda\mu}
    \right),
\end{split}
\end{gather}
and $H_{\psi^*\psi^*} = H_{\psi\psi},~ H_{\psi^*\psi} = H_{\psi\psi^*}$.
To derive them, we have used the relations 
$
    d_\lambda d_\mu = \sum_\nu N_{\lambda\mu}^\nu d_\nu = \sum_\nu N_{\lambda\nu}^\mu d_\nu
$
and
$
    \sum_{\lambda,\nu} N_{\lambda\nu}^\mu d_\lambda d_\nu = D^2 d_\mu.
$
While  the $q$-deformed quadratic Casimir invariant~\cite{Begin:1992rt, Bonatsos:1999xj} was employed in the previous works for $N=2,~3$~\cite{Hayata:2023puo,Hayata:2023bgh}, this work employs the continuum definition~\cite{Gross:1993hu,Zache:2023dko} expressed as
$
    C_2(\lambda) 
    = 
    \frac{1}{2}\left(nN + \tilde{C}_2(\lambda) - \frac{n^2}{N}\right),
    ~
    \tilde{C}_2(\lambda) 
    = 
    \sum_{j=1}^{N-1} \lambda_j (\lambda_j -2j+1),
$
where $n = |\lambda| = \sum_{j=1}^{N-1} \lambda_j$.
This choice has the advantage of being close to the continuum limit in terms of gauge group.

For each $N$ and $k$, we determine the critical coupling $1/g_\mathrm{c}^2$ as the point where the negative eigenvalue of the matrix \eqref{eq:Hessian} appears (or disappears) by varying the coupling from the weak-coupling (or strong-coupling) regime.
Note that there are always two zero modes, coming from the contribution at $k=0$.

\begin{figure}[t]
    \centering
    \includegraphics[width=0.9\linewidth]{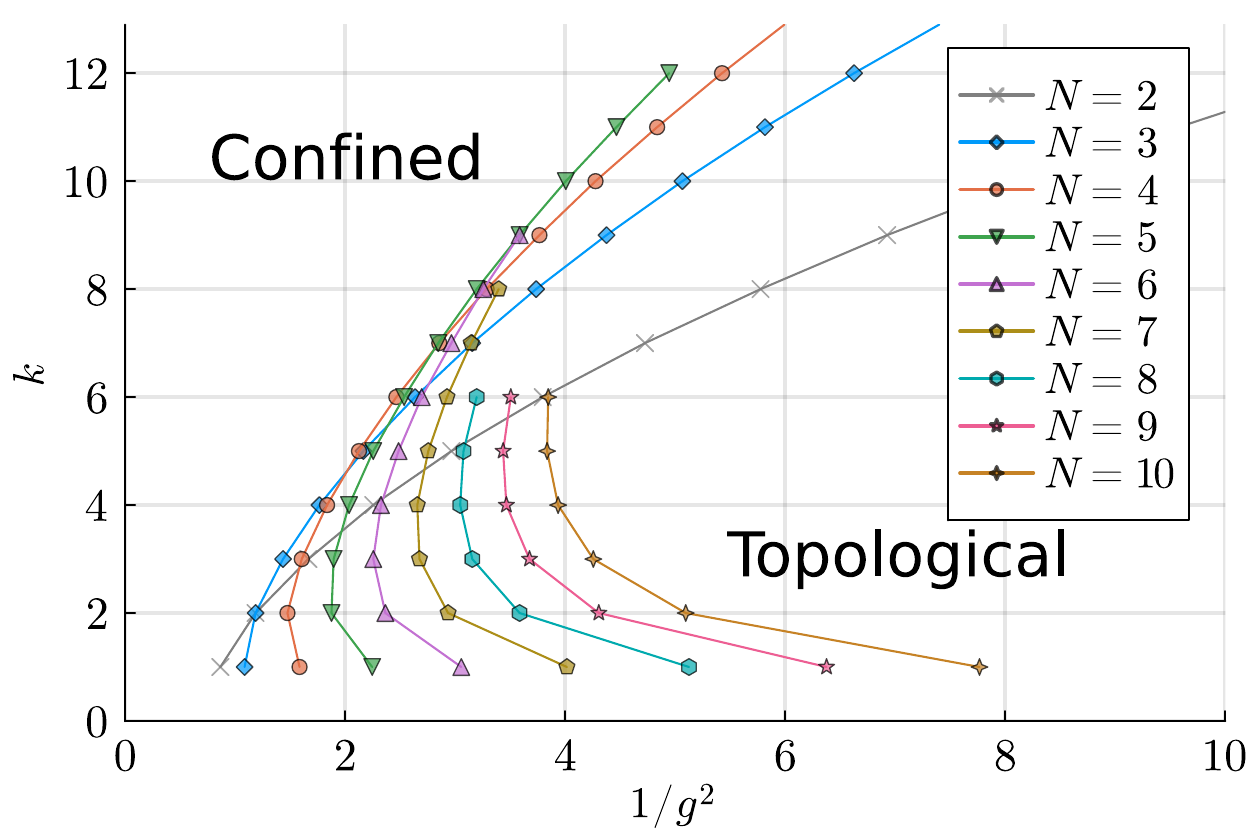}
    \caption{Phase structure in terms of the inverse square of YM coupling $1/g^2$ and the cutoff $k$.}
    \label{fig:phase_boundary}
\end{figure}

\begin{figure}[t]
    \centering
    \includegraphics[width=0.9\linewidth]{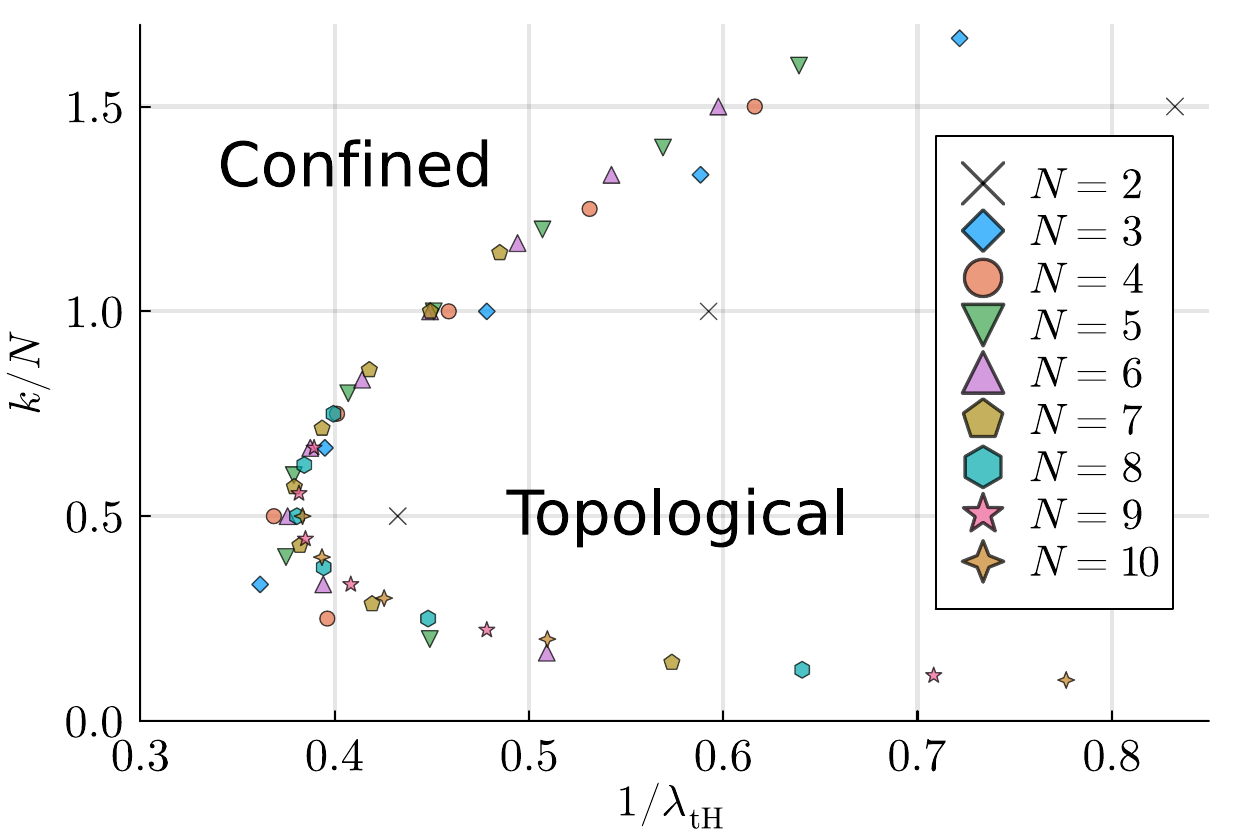}
    \caption{Phase structure in terms of the inverse 't~Hooft coupling $1/\lambda_\mathrm{tH}$ and $k/N$.}
    \label{fig:phase_boundary_thooft}
\end{figure}

\section{Phase diagram at large $N$}

\begin{figure}[th]
    \centering
    \includegraphics[width=0.9\linewidth]{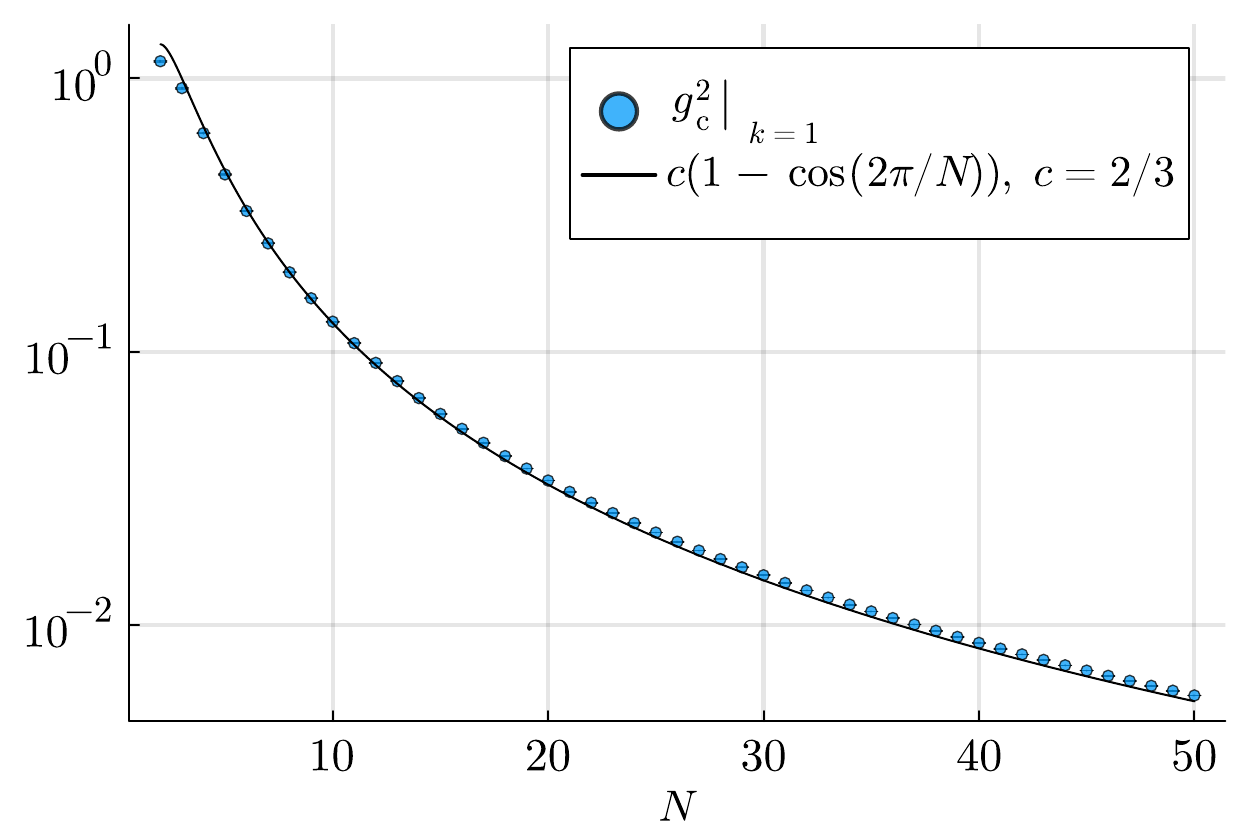}
    \caption{The critical coupling $g_\mathrm{c}^2$ at $k = 1$ for various $N = 2,~\cdots,~50$. The solid line represents a guideline $g_\mathrm{c}^2(N)|_{k=1} = \frac{2}{3}\qty(1-\cos\qty(\frac{2\pi}{N}))$.}
    \label{fig:gc^2_k1_vs_N}
\end{figure}

\begin{figure}[th]
    \centering
    \includegraphics[width=0.9\linewidth]{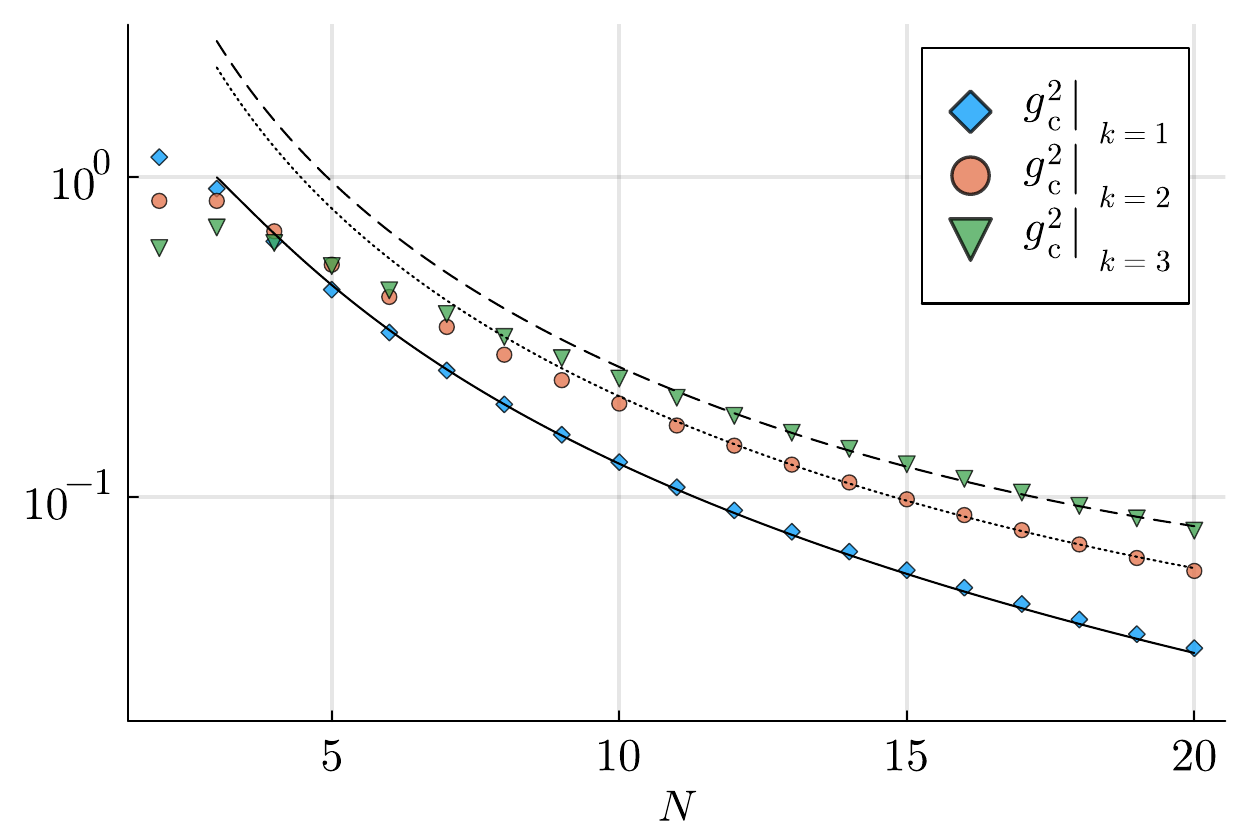}
    \caption{The critical coupling $g_\mathrm{c}^2$ at $k = 1,2,3$ for various $N = 2,~\cdots,~20$.
    The dotted and dashed lines represent the fitted results, while the solid line is a guideline $g_\mathrm{c}^2(N)|_{k=1} = \frac{2}{3}\qty(1-\cos\qty(\frac{2\pi}{N}))$.
    }
    \label{fig:gc^2_k1-3_vs_N}
\end{figure}

Figure~\ref{fig:phase_boundary} shows the phase structure of $\SU(N)_k$ YM in terms of the inverse square of the YM coupling $1/g^2$ and the cutoff $k$. 
Each point corresponds to the critical coupling $1/g_\mathrm{c}^2$, determining the phase boundary at which the string-net condensed state (de)stabilizes in the mean-field calculation.
The phase boundary obtained for $N=3$ shows qualitative agreement with the results of Ref.~\cite{Hayata:2023bgh}, while exhibiting a more extensive confined phase due to the definition of $C_2(\lambda)$.
Moreover, Fig.~\ref{fig:phase_boundary_thooft} represents the same phase structure in terms of the inverse of the 't~Hooft coupling $1/\lambda_\mathrm{tH}$ and $k/N$. 
These figures reveal a couple of notable features. 
First, the phase boundary is drawn nontrivially in the phase diagram:
At $k/N \lesssim 0.5$, the confined phase becomes broader as $N$ increases.
In contrast, for $k/N \gtrsim 0.5$, the critical coupling increases monotonically as a function of cutoff.
This feature strongly suggests that the continuum limit of the $\SU(N)_k$ theory requires a delicate scaling of parameters, otherwise one would experience a bulk transition and intrude the topological phase.
Such a reentrant structure cannot be captured in previous studies restricted to $N=2$ and $N=3$.
Second, the critical 't~Hooft couplings align with a specific band region, except for $N=2$. 
More precisely, $1/\lambda_\mathrm{tH}$ for fixed $k$ lies along with a smooth curve, particularly at large $N$. 
This implies that $N=2$ is not large from the perspective of distinguishing the confined and topological phases at the mean-field level, in contrast with the mass spectra obtained by lattice Monte Carlo simulation~\cite{Teper:1998te}.

The broad confined region at $k=1$ can be interpreted as follows:
The allowed irreducible representations at $k=1$ are labeled by Young diagrams of the form $\lambda^{(n)} = (n,0,~\cdots,~0)$ with $n = 0,~\cdots,~N-1$ and satisfy the fusion rules $\lambda^{(n)} \times \lambda^{(m)} = \lambda^{({n+m}\mod N)}$.
In other words, the $\SU(N)_{k=1}$ theory in the weak-coupling limit realizes an abelian topological order: All line operators correspond to abelian anyons with quantum dimension one, and the resulting ground state is described by the $\mathbb{Z}_N$ topological quantum field theory~\cite{Moore:1988qv,Hsin:2018vcg}.
In this sense, it effectively has a structure very similar to that of a $\mathbb{Z}_N$ gauge theory.
A naive large-$N$ limit of such a $\mathbb{Z}_N$ theory leads to a $\mathrm{U}(1)$ theory, which admits only a confining vacuum in $(2+1)$ dimensions~\cite{Polyakov:1975rs,Polyakov:1976fu}.
As discussed below, this argument is supported by quantitative agreement with lattice simulations.

Somewhat unexpectedly, we also find that, for $k > 1$, the confined phase extends in a nontrivial way as $N$ increases.
For $k\ge2$, nonabelian anyons with quantum dimension larger than one enter the Hilbert space and become more dominant for larger $N$.
Moreover, for $N \ge 3$, nontrivial fusion multiplicities $N_{\lambda\mu}^\nu \ge 2$ start to appear at $k=3$.
These arguments might suggest that the abelian structure observed at $k=1$ does not persist for larger $k$.
In contrast, our results indicate that an extensive confined region survives for $k \ge 2$.
Of course, one must take into account the additional contribution from the electric field operator appearing in the first term of Eq.~\eqref{eq:Hamiltonian_MF}, which breaks the topological order away from the weak coupling regime.
The phase structure is determined by a competition between the electric field and the Wilson loop operators, achieved through the minimization of the ground-state energy. 
In the small-$k$ regime, the observed phenomenon is therefore a direct manifestation of these nontrivial dynamical effects.

The absence of the level-rank duality in the phase boundaries can also be examined from a similar perspective.
In the Wess–Zumino–Novikov–Witten models, one finds a level-rank duality that relates certain algebraic data of $\SU(N)_k$ to those of $\SU(k)_N$, such as the fusion rules or quantum dimensions~\cite{Nakanishi:1990hj}.
In the weak-coupling limit, the second term of the mean-field Hamiltonian density~\eqref{eq:Hamiltonian_MF}, which consists of such quantities, dominates and the duality is manifested.
On the other hand, our quadratic Casimir invariant does not respect this duality. 
As a result, the phase boundaries are not symmetric under $N \leftrightarrow k$. 

We discuss more quantitatively the expansion of the confined phase at small cutoff $k$.
Figure~\ref{fig:gc^2_k1_vs_N} shows the $N$ dependence of the critical coupling for $k=1$.
The damping behavior remarkably agrees with the solid black line representing $g_\mathrm{c}^2(N)|_{k=1} = \frac{2}{3}\qty(1-\cos\qty(\frac{2\pi}{N}))$ observed in the lattice Monte Carlo simulation~\cite{Bhanot:1980pc}, which indicates $g_\mathrm{c}^2 \sim N^{-2}$ at large $N$~\cite{Horn:1979fy}. 
Moreover, Fig.~\ref{fig:gc^2_k1-3_vs_N} shows the $N$ dependence of the critical coupling for $k=1,2,3$.
As in Fig.~\ref{fig:gc^2_k1_vs_N}, the solid line represents $g_\mathrm{c}^2(N)|_{k=1}$. 
The dotted and dashed lines are the fitted results by a similar ansatz $g_\mathrm{c}^2(N)|_{k}=(c^{(k)}_2 N^2 + c^{(k)}_4N^4)^{-1}$, and we obtained $c_2^{(k=2)}=0.0506(5),~c_4^{(k=2)}=-2.3(2)\times10^{-5}$ and $c_2^{(k=3)}=0.0419(6),~c_4^{(k=3)}=-2.8(3)\times10^{-5}$, respectively.
As in the case of $k=1$, we observe that the confined phase expands nontrivially as $N$ increases.

\begin{figure}[t]
    \centering
    \includegraphics[width=0.9\linewidth]{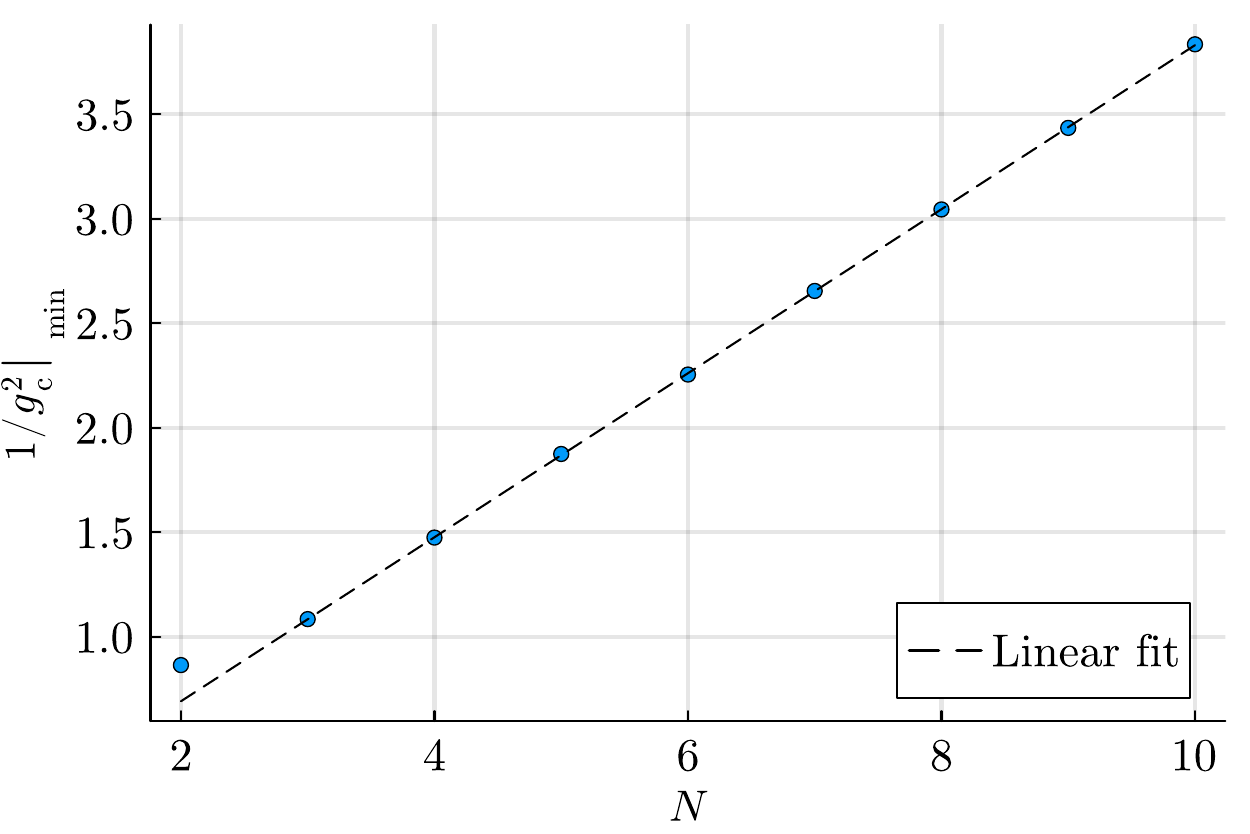}
    \caption{
        Minimum of the inverse critical coupling for $N = 2,~\cdots,~10$.
        The dashed line represents the fitted result of a linear function.
    }
    \label{fig:g2inv_min_vs_N}
\end{figure}

Figure~\ref{fig:g2inv_min_vs_N} shows the $N$ dependence of the minimum of the critical YM coupling $1/g_\mathrm{c}^2(N)|_\mathrm{min}\equiv \underset{k}{\mathrm{min}}\big(1/g_\mathrm{c}^2(N,k)\big)$ for $N = 2,\cdots,10$.
For $N\ge 3$, the data are well fitted by a linear function, $1/g_\mathrm{c}^2(N)|_\mathrm{min} = c_1 N + c_0$ with $c_1 = 0.3918(8)$ and $c_0 = -0.090(5)$. 
This behavior implies that the minimum of the inverse critical 't~Hooft coupling, $(1/\lambda_\mathrm{tH})|_\mathrm{min} = 1/(g_\mathrm{c}^2N)|_\mathrm{min}$ is almost independent of the value of $N$.
Moreover, Tab.~\ref{tab:kext} presents the corresponding cutoff, $k_\mathrm{ext}(N)\coloneqq k(1/g_\mathrm{c}^2(N)|_\mathrm{min})$.
Taken together, this analysis indicates that the extremal point $(1/\lambda_\mathrm{tH}, k/N) \approx (0.4, 0.5)$ plays an important role as a reference point for the parameters when taking the continuum limit $k\to\infty$ and $1/g^2 \to \infty$ with fixed $N$.

\begin{table}[t]
    \centering
    \caption{List of $k_\mathrm{ext}$ for each $N$.}
    \begin{tabular}{|wc{1.0cm}|wc{0.65cm}|wc{0.65cm}|wc{0.65cm}|wc{0.65cm}|wc{0.65cm}|wc{0.65cm}|wc{0.65cm}|wc{0.65cm}|wc{0.65cm}|}
        \hline
         $N$ & 
         2&3&4&5&6&7&8&9&10
         \\
         \hline
         $k_\mathrm{ext}$& 
         1& 1& 2& 2& 3& 4& 4& 5& 5
         \\
         \hline
         $k_\mathrm{ext}/N$ &
         0.5& 0.33& 0.5& 0.4& 0.5& 0.57& 0.5& 0.56 & 0.5
         \\
         \hline
    \end{tabular}
    \label{tab:kext}
\end{table}

In the region $k/N \gtrsim 0.5$, where the critical couplings grow monotonically as a function of $k$, we fit the data for the region $k \ge k_\mathrm{ext}$ using 
\begin{equation}
    \frac{k(\lambda_\mathrm{tH})}{N} 
    = 
    A\qty(\frac{1}{\lambda_\mathrm{tH}} - \frac{1}{\lambda_0})^B + \frac{k_\mathrm{ext}}{N} + C,
    \label{eq:fitting_ansatz}
\end{equation}
with fitting parameters $A$, $1/\lambda_0$, $B$, and $C$.
Note that an essentially equivalent fitting ansatz was employed in Ref.~\cite{Hayata:2023bgh} for the case $N=3$.
The results of our best fits are summarized in Tab.~\ref{tab:fitting_result}; see the Supplemental Material for further details of the fitting results, where we also show the fitting results of the critical curves in the region $k/N \lesssim 0.5$.
Except for $N=2$, the fitted parameters $A, 1/\lambda_0, B$ show good agreement with one another.
Moreover, the value of $1/\lambda_0$ is also consistent with $c_1 = 0.3918(8)$ obtained above.
The fitting result suggests the possibility of a continuum limit in terms of $\lambda_\mathrm{tH}$ and $k/N$, even at large $N$.

\begin{table}[t]
    \centering
    \caption{Fitting results by Eq.~\eqref{eq:fitting_ansatz}.}
    \begin{tabular}{|cc|cccc|}
        \hline
        $N$ &$k_\mathrm{ext}$ &$A$        &$1/\lambda_0$  &$B$        &$C$      \\ \hline
        2   &1      &2.16(3)    &0.450(3)       &0.597(7)   &-0.20(2) \\
        3   &1      &2.61(3)    &0.358(3)       &0.687(6)   &-0.09(4) \\
        4   &2      &2.69(3)    &0.356(3)       &0.600(6)   &-0.19(3) \\
        5   &2      &2.59(7)    &0.375(5)       &0.647(33)  &0.11(4)  \\
        6   &3      &2.54(4)    &0.376(14)      &0.623(101) &0.00(19) \\
        7   &4      &2.68(7)    &0.379(0)       &0.688(19)  &-0.00(1) \\ \hline
    \end{tabular}
    \label{tab:fitting_result}
\end{table}

\section{Summary and outlook}

We have analyzed the phase structure of the $q$-deformed $\SU(N)$ YM theory in $(2+1)$ dimensions using mean-field theory.
By computing the Hessian around the string-net condensed state, we determined the critical lines separating confined and topological phases for $2\le N\le 10$ and a wide range of $k$.
When plotted in terms of $1/\lambda_\mathrm{tH}$ and $k/N$, the critical lines for $N\ge 3$ collapse onto a universal curve, indicating a well-defined large-$N$ limit and suggesting that the continuum limit is more nontrivial than one might anticipate.
We also showed that for $k=1$ the confined phase broadens with increasing $N$ in a way consistent with the large-$N$ limit of $\mathbb{Z}_N$ gauge theory, 
while for $k/N\gtrsim 0.5$ the topological phase persists at weak coupling even as $N\to\infty$.

There are several directions for future work.
First, analyses beyond the mean-field treatment will be essential. 
In particular, determining the order of the phase transition is an important open problem.
In the present work, we assumed a second-order phase transition in the stability analysis, and the properties of the phase transition line are not determined for general $N$ even at the mean-field level.
Second, it would be valuable to relate the phase diagram in the ($\lambda_\mathrm{tH},k/N$) plane more directly to renowned continuum large-$N$ results obtained in the path-integral formalism~\cite{Gross:1980he,Wadia:1980cp,Douglas:1993iia,Aganagic:2004js,Naculich:2007nc,Arsiwalla:2005jb,Narayanan:2006rf}. 
It would also be interesting to make direct comparisons with lattice simulations of $(2+1)$-dimensional $\SU(N)$ pure gauge theory~\cite{Bursa:2006tm}.
Third, the algebraic data computed here (fusion rules, modular $S$-matrices, and quantum dimensions) are useful in quantum-simulation platforms, guiding the design of analog or digital quantum simulations of large-$N$ gauge theories.
Finally, extending our analysis to include matter fields, or alternative truncation schemes, would clarify how universal the topological order at large $N$ is in more general settings.
Specifically, choices of the quadratic Casimir invariant would alter the phase structure, particularly in the regions with a small cutoff $k$, opening an intriguing future direction for engineering the continuum limit.

\section{Acknowledgments}
\begin{acknowledgments}
T.~H. was supported by JSPS KAKENHI Grants No.~JP24K00630 and No.~JP25K01002.
T.~H. is also grateful for the funding received from JST COI-NEXT (Grant No.~ JPMJPF2221).  
Y.~H. was supported by JSPS KAKENHI Grants No.~JP24H00975, No.~JP24K00630, No.~JP25K01002, and by JST, CREST Grant Number JPMJCR24I3.
H.~W. was supported by JSPS KAKENHI Grant No.~JP24K00630.
\end{acknowledgments}

\bibliographystyle{utphys}
\bibliography{ref.bib}

\end{document}

\begin{titlepage}

\title{SUPPLEMENTAL MATERIAL: \\
Phases of the $q$-deformed $\SU(N)$ Yang-Mills theory at large $N$}

\author{Tomoya Hayata}
\email{hayata@keio.jp}
\affiliation{Department of Physics, Keio University, 4-1-1 Hiyoshi, Yokohama, Kanagawa 223-8521, Japan}
\affiliation{RIKEN Center for Interdisciplinary Theoretical and Mathematical Sciences
(iTHEMS), RIKEN, Wako 351-0198, Japan}
\affiliation{International Center for Elementary Particle Physics and The University of Tokyo, 7-3-1 Hongo, Bunkyo-ku, Tokyo 113-0033, Japan}

\author{Yoshimasa Hidaka}
\email{yoshimasa.hidaka@yukawa.kyoto-u.ac.jp}
\affiliation{Yukawa Institute for Theoretical Physics, Kyoto University, Kitashirakawa Oiwakecho, Sakyo-ku, Kyoto 606-8502, Japan}
\affiliation{RIKEN Center for Interdisciplinary Theoretical and Mathematical Sciences
(iTHEMS), RIKEN, Wako 351-0198, Japan}

\author{Hiromasa Watanabe}
\email{hiromasa.watanabe@keio.jp}
\affiliation{Department of Physics, Keio University, 4-1-1 Hiyoshi, Yokohama, Kanagawa 223-8521, Japan}
\affiliation{Research and Education Center for Natural Sciences, Keio University, 4-1-1 Hiyoshi, Yokohama, Kanagawa 223-8521, Japan}

\maketitle
\end{titlepage}

\section{Computation of algebraic data}

\subsection{Modular $S$-matrix}
\label{sec:modularS}
Irreducible representations of $\SU(N)_k$ can be characterized by partitions\footnote{
Recall that an $\SU(N)$ partition $\lambda$ can be expressed by a non-increasing non-negative integer series $\lambda_1\ge \lambda_2 \ge \cdots \ge \lambda_{N-1} \ge \lambda_N = 0$.
}, which have a graphical representation in terms of the Young diagrams with boxes of height (: rows) $N-1$ and width (: columns) $k$.
For given partitions $\lambda,\mu$, the modular $S$-matrix of $\SU(N)_k$ is given by the Kac-Peterson formula~\cite{Kac:1984mq} as 
\begin{equation}
    S_{\lambda\mu}
    =
    \frac{\ee^{\ii \pi N(N-1)/4}}{\sqrt{N(k+N)^{N-1}}}
    \sum_{w\in W}(-1)^{\ell(w)}\ee^{-\frac{2\pi\ii}{k+N}\ev{\mu+\rho,w(\lambda+\rho)}},
    \label{eq:S-matrix}
\end{equation}
where $W$ is the Weyl group, which turns out to be the $S_N$ permutation group, and $\ell(w)$ is the length of the element $w\in W$.
Here, 
$\rho=\sum_{\alpha \in \Delta^+} \alpha$ is the Weyl vector, which is a sum of positive root vectors whose concrete form is $\rho = \qty(N-1, N-2,~\cdots,~0)$, 
and $\ev{\cdot,\cdot}$ is the $\SU(N)$ inner product:
\begin{equation}
    \ev{u,v} = \sum_{j=1}^N u_j v_j - \frac{1}{N}\qty(\sum_{j=1}^N u_j)\qty(\sum_{j=1}^N v_j),
\end{equation}
in which the trace part is subtracted to drop an unnecessary $\mathrm{U(1)}$ component.

\subsection{An efficient approach to computing $S$-matrices}
If one naively evaluates \eqref{eq:S-matrix}, it requires a computational cost of $O(N!)$ because of $|S_N| = N!$.
A remarkable fact for reducing the cost is that the modular $S$-matrix has an equivalent determinant formula that can be expressed in terms of various symmetric polynomials.
For the details of symmetric polynomials, see e.g., Ref.~\cite{10.1093/oso/9780198534891.001.0001}.

Let us introduce $x_j$ using the $q$-deformation parameter 
\begin{align}
&        q \coloneqq \ee^{\frac{2\pi\ii}{k+N}}, \label{eq:q_number}
    \\
&    x_j 
    \coloneqq
    q^{-\qty(\mu_j + N-j - \frac{1}{N}\sum_{j=1}^N  (\mu_j + N-j))}
    =
    q^{-\qty(\tilde\mu_j + N-j)} ,
\end{align}
with $\tilde\mu_j = \mu_j - \frac{1}{N}\sum_{j=1}^N  (\mu_j + N-j)$.
Note that these variables satisfy 
\begin{equation}
    \prod_{j=1}^N x_j = 1.
    \label{eq:centering_x}
\end{equation}
Then, by denoting $X = \qty{x_1,\cdots,x_N}$, the expression in Eq.~\eqref{eq:S-matrix} can be written as
\begin{equation}
\begin{split}
        S_{\lambda\mu}
    &\propto
    \det_{1\le i,j \le N} \qty(x_j^{\tilde\lambda_i+N-i})
    \\
    &=
    \det_{1\le i,j \le N} \qty(x_j^{\lambda_i+N-i})
    =
    s_{\lambda}(X)\cdot \Delta(X)
    ,
\end{split}
\end{equation}
up to a numerical factor.
Here, we have introduced the Schur polynomial 
\begin{equation}
    s_{\lambda}(X)
    =
    \frac{\displaystyle\det_{1\le i,j\le N}\qty(x_j^{\lambda_i+N-i})}{\displaystyle\det_{1\le i,j\le N}\qty(x_j^{N-i})},
\end{equation}
and the Vandermonde determinant
\begin{equation}
    \Delta(X) 
    =
    \det_{1\le i,j\le N}\qty(x_i^{N-j})
    =
    \prod_{i<j} (x_i - x_j).
\end{equation}
At this stage, we have already reduced the computational cost to $O(N^3)$, coming from the evaluation of the determinant by the LU decomposition.

Further reductions can be achieved as follows.
By using the power-sum symmetric polynomials 
\begin{equation}
    p_m(X) = \sum_{j=1}^N x_j^m,
\end{equation}
the complete homogeneous symmetric polynomials can be obtained through Newton's identity
\begin{equation}
    r h_r(X) = \sum_{m=1}^r p_m(X) h_{r-m}(X),
\end{equation}
with conditions $h_0(X) = 1,~h_{r<0}(X) = 0$.
The index runs at most up to
\begin{equation}
    r_\mathrm{max} 
    =
    \max_{i,j}(\lambda_i-i+j)
    =
    \lambda_1 - 1 + \ell(\lambda).
\end{equation}
Using the fact $\lambda_1 \le k$, it is sufficient to compute $\qty{h_r}$ with $r=1,~\cdots,~R = k + N - 2$ for each $\mu$, or equivalently, $X$.

These nice symmetric polynomials enable us to compute the Schur polynomial through the Jacobi--Trudi formula, expressed as
\begin{equation}
    s_{\lambda}(X) = \det_{1\le i,j \le \ell(\lambda)} \qty(h_{\lambda_i-i+j}(X)).
\end{equation}
We have succeeded in reducing the cost of calculating this determinant to $O(\ell(\lambda)^3)$ with $\ell(\lambda) \le N - 1$.
Moreover, once we compute $X$ and $h_r$ for each $\mu$, we can reuse them for Schur polynomials of any irreducible representation $\lambda$, which enables further efficiency gains for obtaining the entire element of the modular $S$-matrix.

\subsection{Properties of $S$-matrix}
The $S$-matrix enjoys several notable properties (see also~\cite{Moore:1988qv}):
\begin{itemize}
    \item Unitarity: $S^\dagger S = 1$, and $\overline{S_{\lambda\mu}} = S_{\lambda\mu}^{-1}$.
    
    \item Crossing symmetry: $\overline{S_{\lambda\mu}} = S_{\bar{\lambda}\mu}$, where $\bar\lambda=(\lambda_1 - \lambda_N, \lambda_1 - \lambda_{N-1},~\cdots,~\lambda_1-\lambda_2)$ is the conjugate of representation $\lambda$.

    \item Charge conjugation $C$ is obtained by $C=S^2$, and $C_{\lambda\mu} = \delta_{\lambda\bar{\mu}}$.

    \item Modular relation: $(ST)^3 = C$ where
    \begin{gather}
        T_{\lambda\lambda} = \exp[2\pi\ii\qty(h_\lambda - \frac{c}{24})],
        \\
        h_\lambda = \frac{\ev{\lambda,\lambda+2\rho}}{2(k+N)},
        \quad
        c = \frac{k \dim \SU(N)}{k+N}.
    \end{gather}

    \item The quantum dimension $d_\lambda$ can be computed as $d_\lambda = S_{\lambda\emptyset}/S_{\emptyset\emptyset}$.
\end{itemize}

The above properties are useful for the consistency check in the numerical implementation of the modular $S$-matrix.

\subsection{Verlinde formula}
For the computation of fusion coefficients through the Verlinde formula, we employ the following strategies to bypass part of the calculation.
\begin{itemize}
    \item Practically, checking $N$-ality~\cite{Rothe:1992nt,Greensite:2003bk} is quite effective in eliminating the zero fusion coefficients from the beginning.
    For $\SU(N)$, the $N$-ality is a genuine quantum number to be preserved, namely 
    \begin{equation}
    \abs{\lambda} + \abs{\mu} - \abs{\nu} = 0\mod N .
    \end{equation}
    Therefore, the number of boxes for irreducible representations to fuse is first checked in the simulation code.
    If they do not satisfy the $N$-ality condition, the corresponding fusion coefficient must vanish, and we can skip the calculation of the Verlinde formula numerically.
    
    \item Fusion coefficients are symmetric under an exchange of representations to fuse, namely, $N_{\lambda\mu}^\nu = N_{\mu\lambda}^\nu$.
    Using this, we can skip half of the computations.
    Note that further reduction can be achieved if one utilizes other symmetries of fusion coefficients with respect to raising/lowering indices and taking conjugates, such as $N_{\lambda\mu}^\nu =  N_{\lambda\bar\nu}^{\bar\mu} = N_{\bar\lambda\bar\mu}^{\bar\nu}$.

    \item If the sum of the first element of the partitions being fused is less than the cutoff $k$, the corresponding fusion coefficient coincides with the undeformed one, i.e., that of $\SU(N)$.
    In other words, the cutoff dependence appears when fusing irreducible representations $\lambda, \mu$ with $\lambda_1 + \mu_1 \ge k$. 
    In this analysis, we update only such cases and reuse the coefficients obtained at small $k$.
\end{itemize}

\subsection{Quantum dimension}
Although the quantum dimension for irreducible representations $d_\lambda$ can be calculated from the corresponding $S$ matrices, we utilize the following formula~\cite{Aganagic:2004js, Naculich:2007nc}
\begin{equation}
    d_\lambda = \prod_{1\le i < j \le N} \frac{[\lambda_i - \lambda_j -i + j]}{[j - i]},
    \label{eq:q_dim_formula}
\end{equation}
where the square bracket gives the $q$-number defined through the $q$-deformation parameter as
\begin{equation}
    %q = \exp\left(\frac{2\pi\mathrm{i}}{k+N}\right),
    %\qquad
    [x] = \frac{q^{x/2} - q^{-x/2}}{q^{1/2} - q^{-1/2}},
\end{equation}
with $q$ given in Eq.~\eqref{eq:q_number}.
Note that this is a $q$-deformed version of the well-known formula computing the dimension of irreducible representations, and therefore, Eq.~\eqref{eq:q_dim_formula} agrees with the standard definition in the Lie algebra by taking the continuum limit for the quantum group, $k\to\infty$.

\begin{figure*}
    \includegraphics[width=0.32\linewidth]{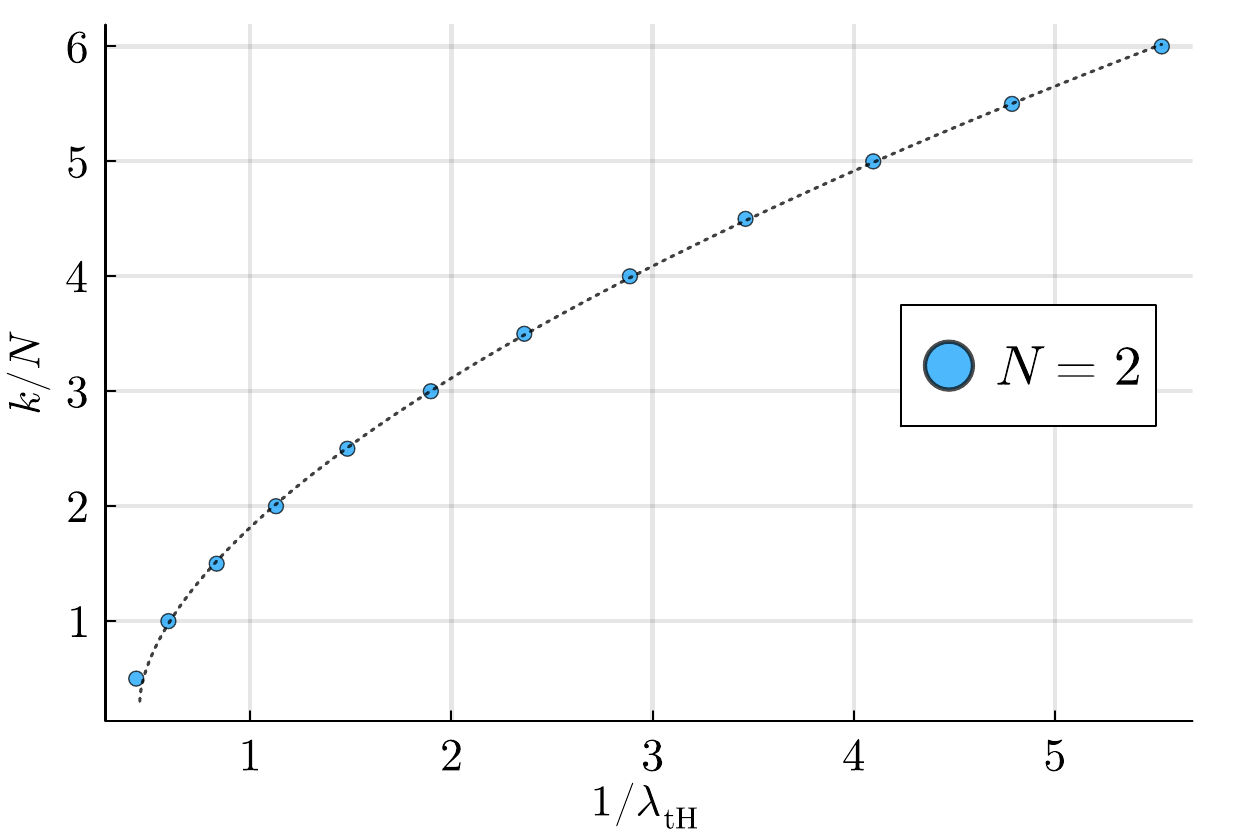}
    \includegraphics[width=0.32\linewidth]{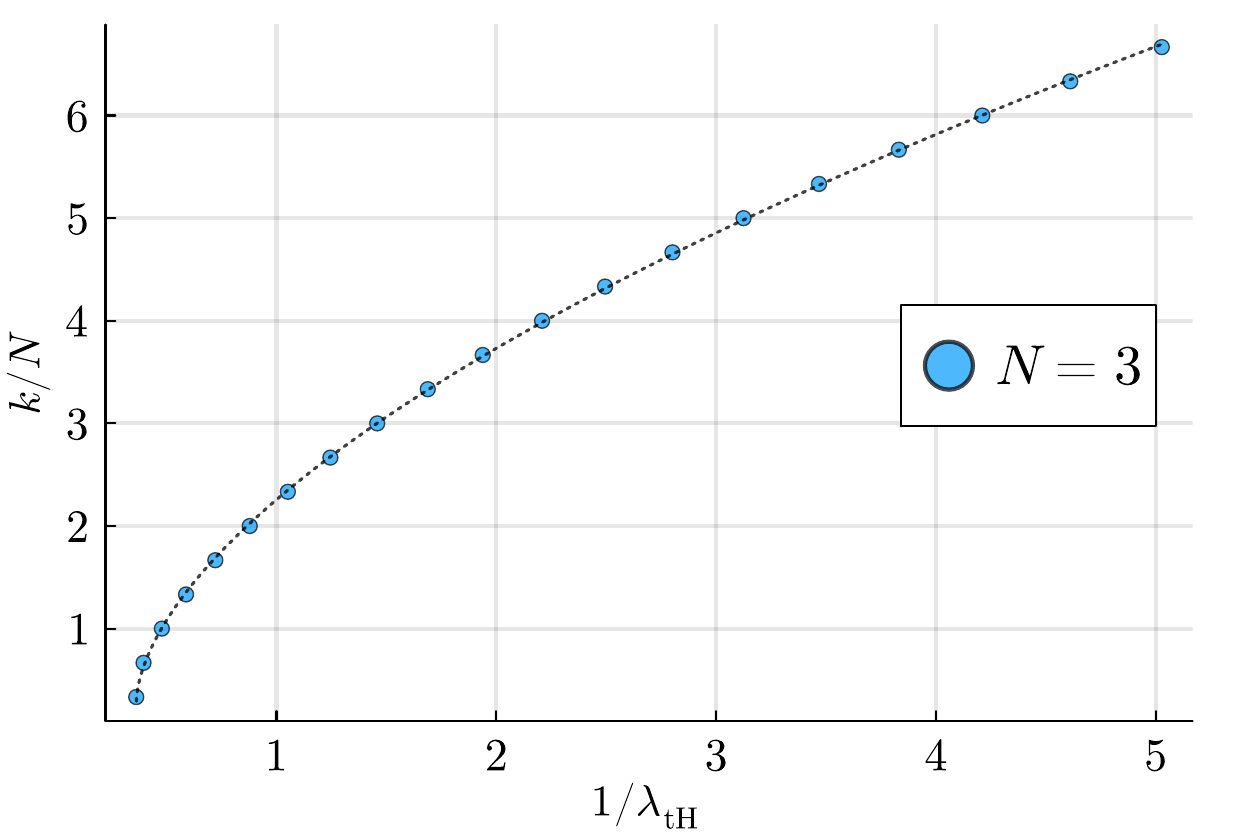}
    \includegraphics[width=0.32\linewidth]{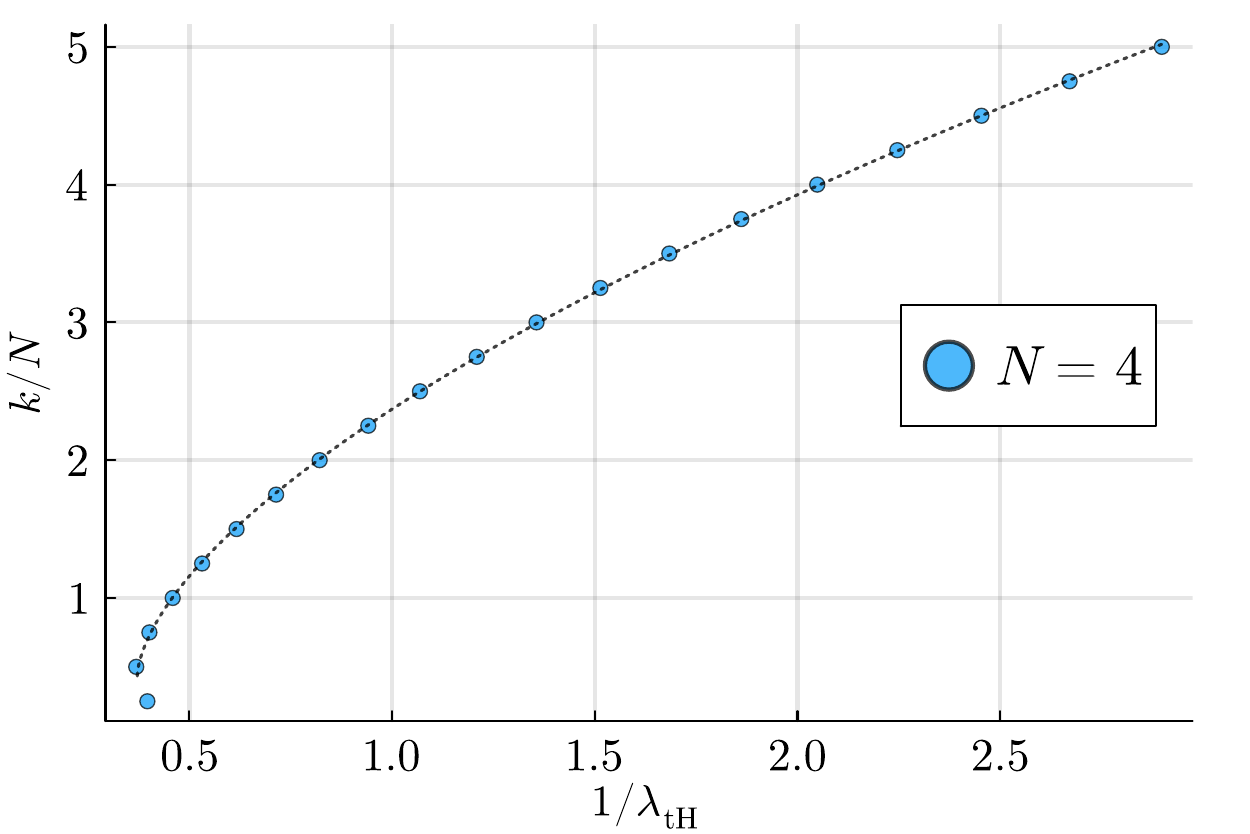}
    \\
    \includegraphics[width=0.32\linewidth]{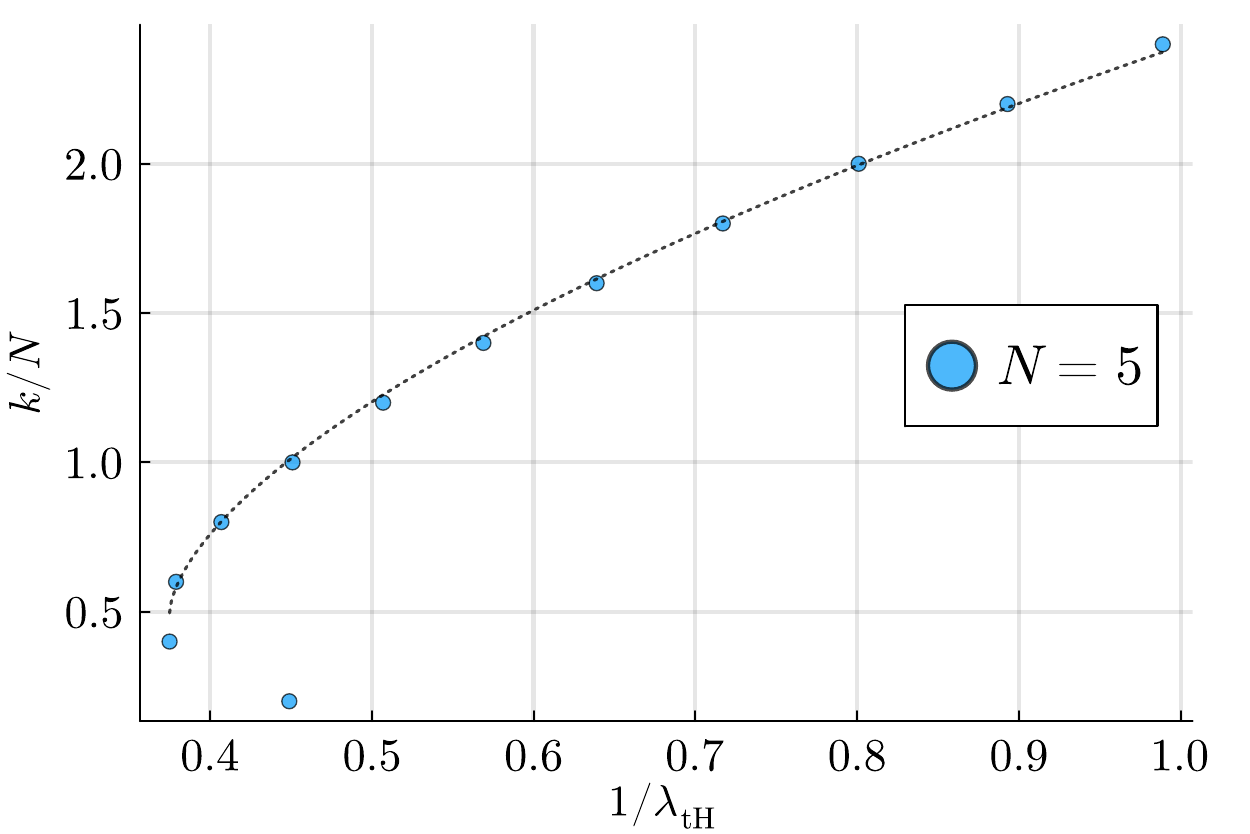}
    \includegraphics[width=0.32\linewidth]{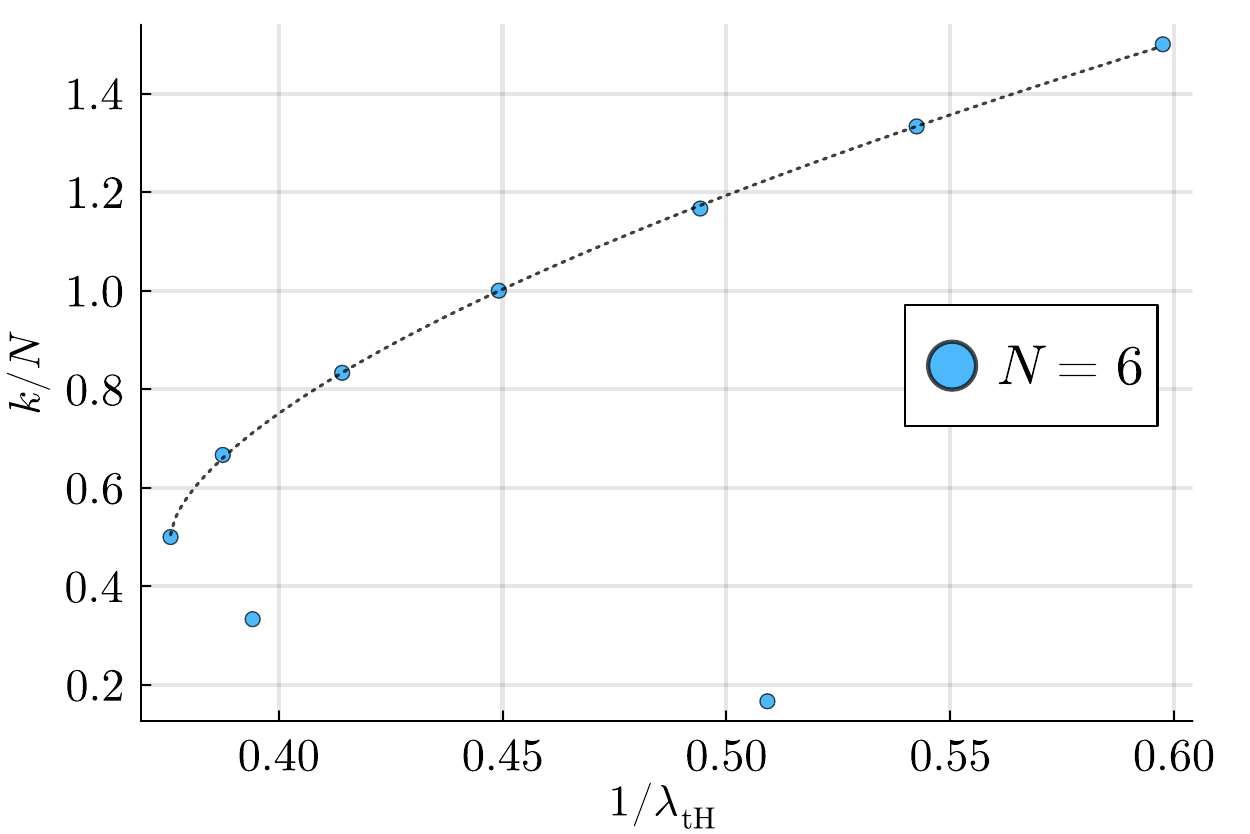}
    \includegraphics[width=0.32\linewidth]{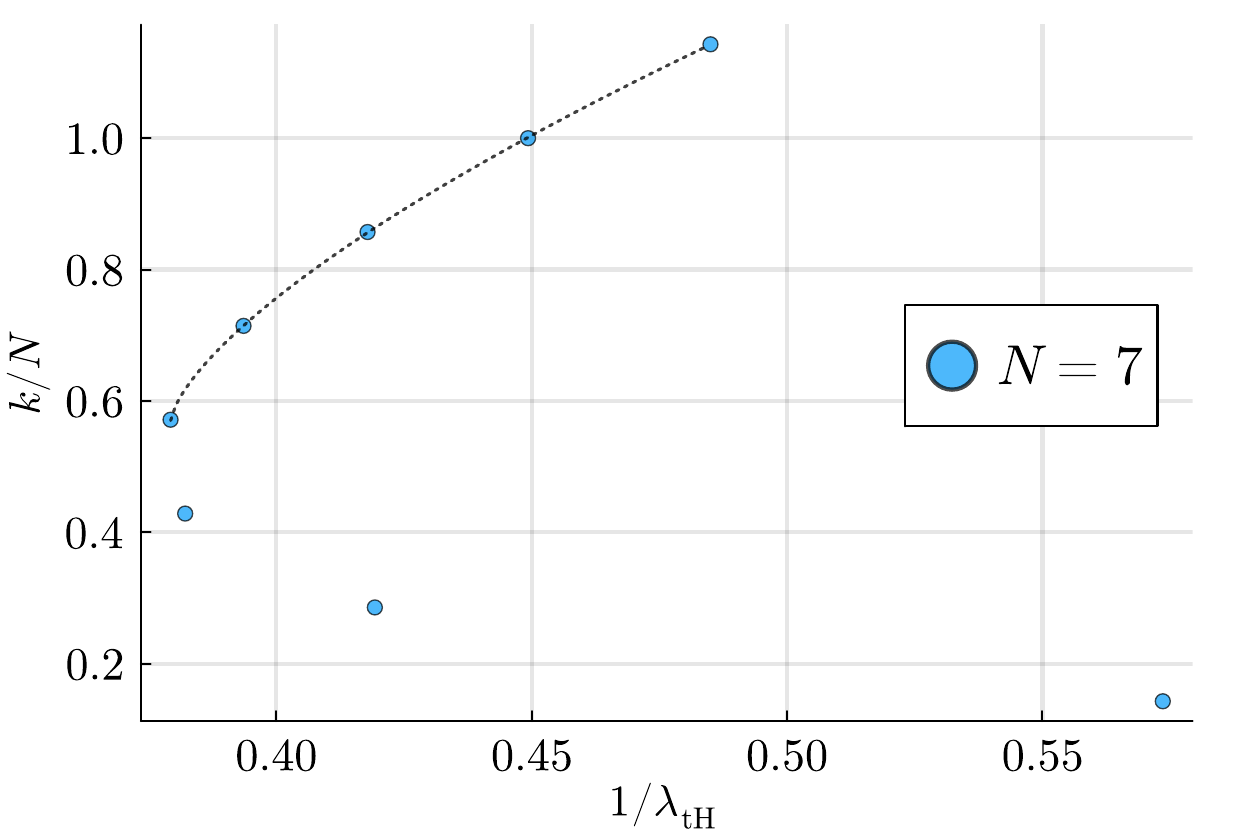}
    \caption{
    Phase diagrams in terms of the inverse 't~Hooft coupling $1/\lambda_\mathrm{tH}$ and the ratio $k/N$.
    The dotted lines represent the fitting results in the region $k \ge k_\mathrm{ext}$ with the ansatz in Eq.~(9) of this Letter.
    }
    \label{fig:lambdainv_vsk/N_upper}
\end{figure*}

\begin{figure*}
    \includegraphics[width=0.32\linewidth]{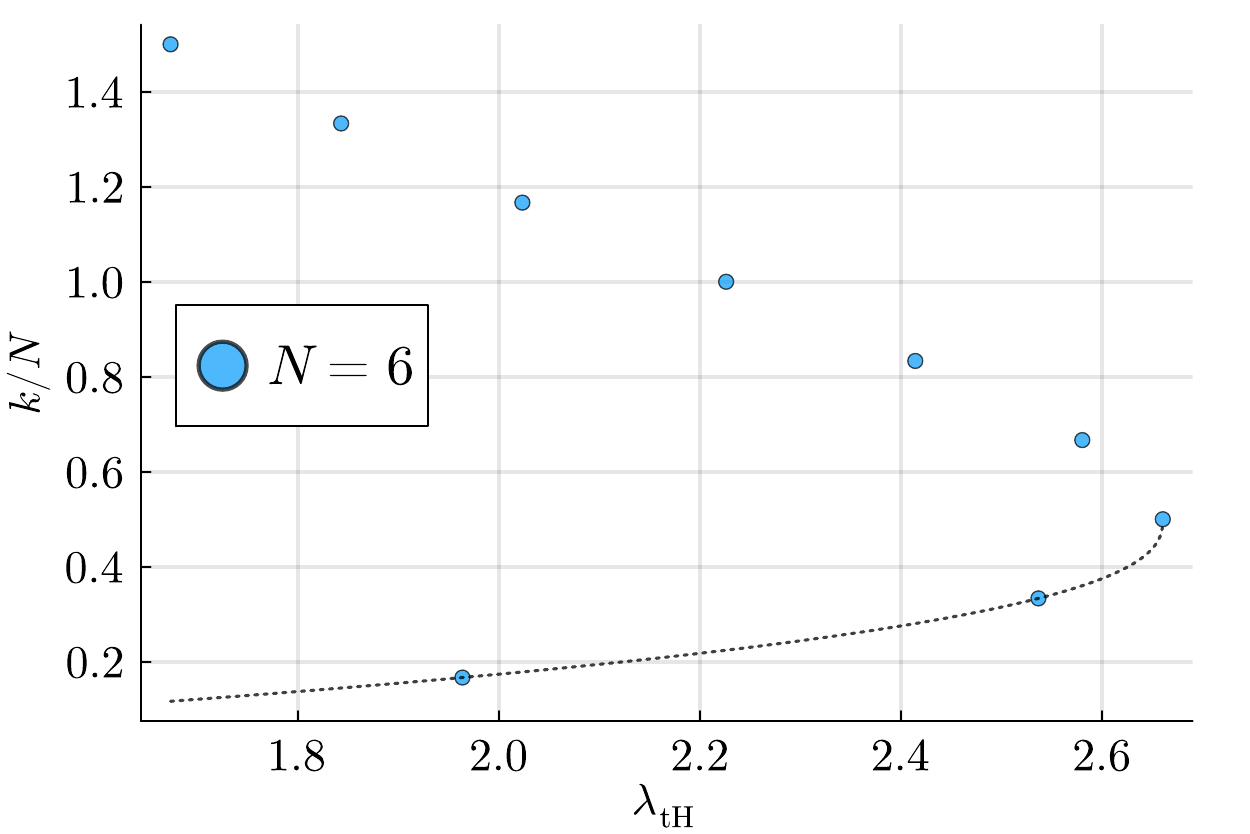}
    \includegraphics[width=0.32\linewidth]{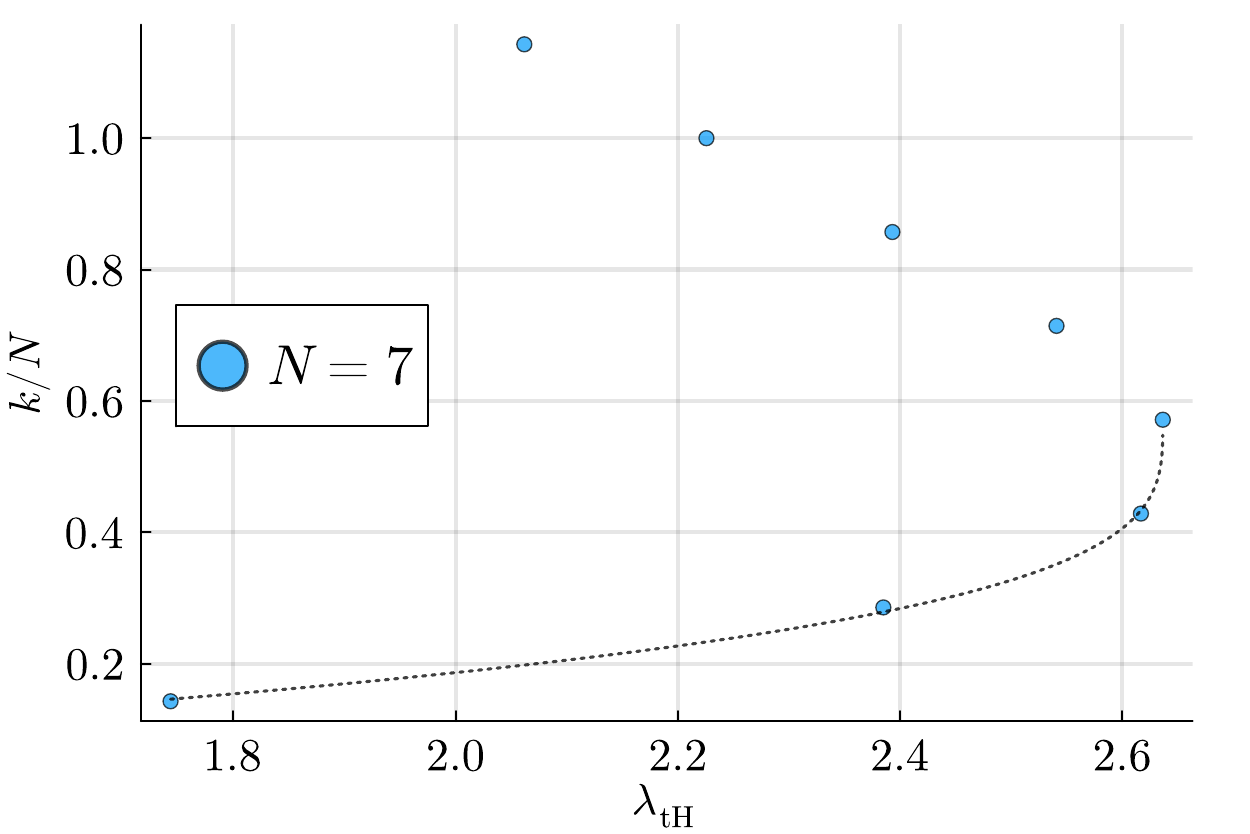}
    \includegraphics[width=0.32\linewidth]{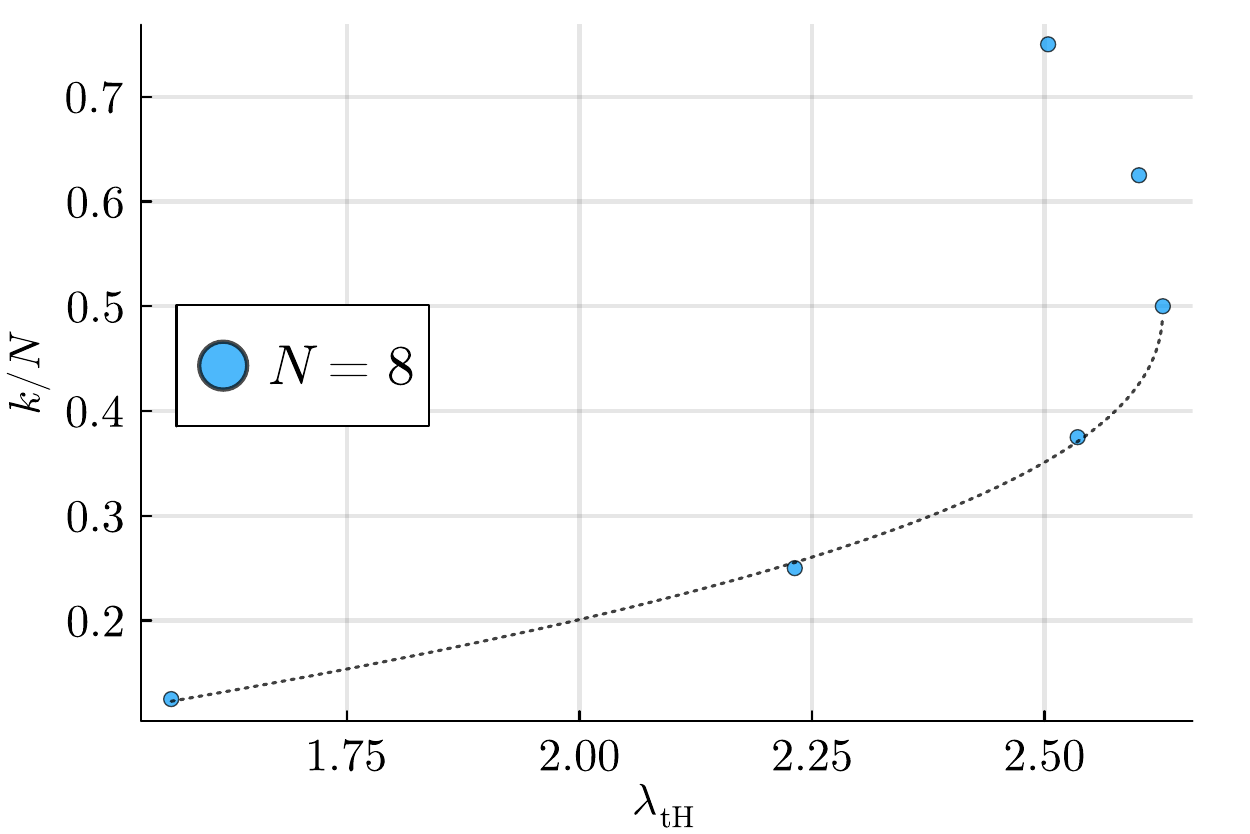}
    \\
    \includegraphics[width=0.32\linewidth]{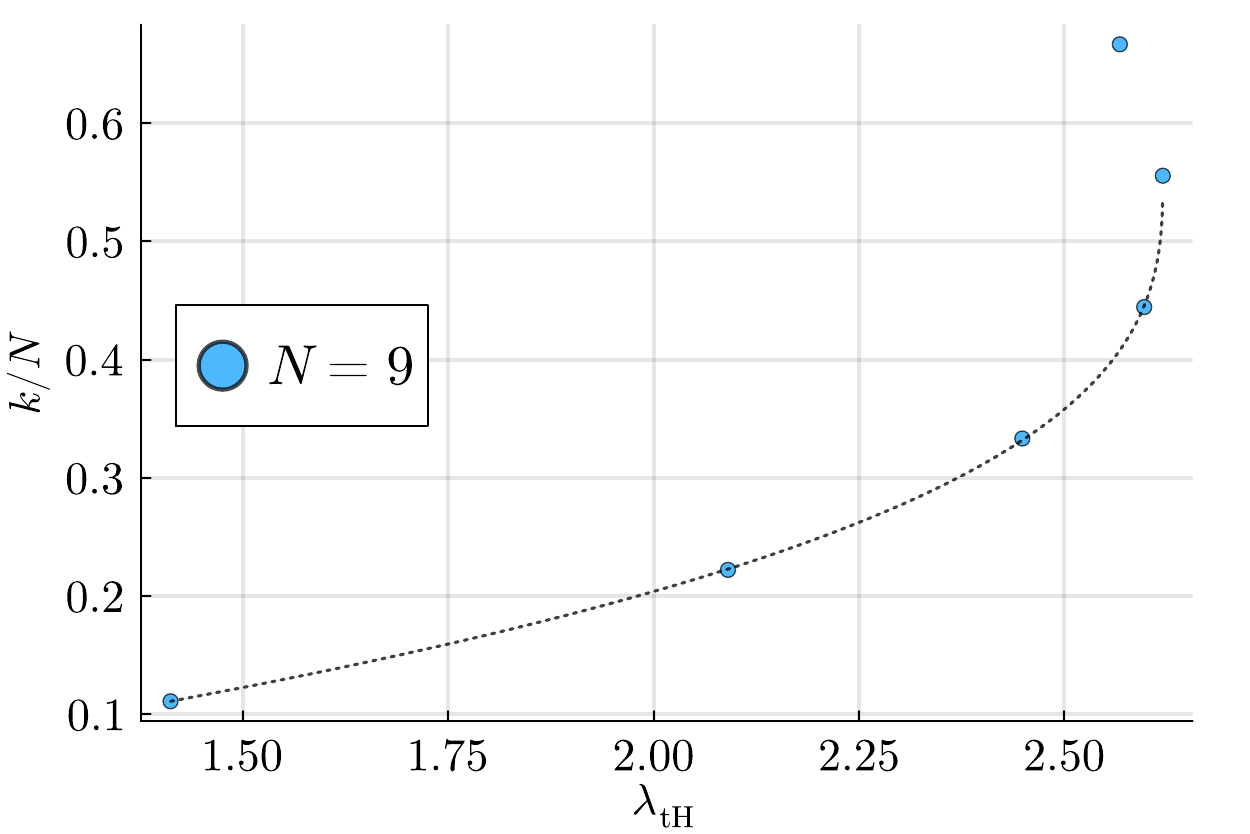}
    \includegraphics[width=0.32\linewidth]{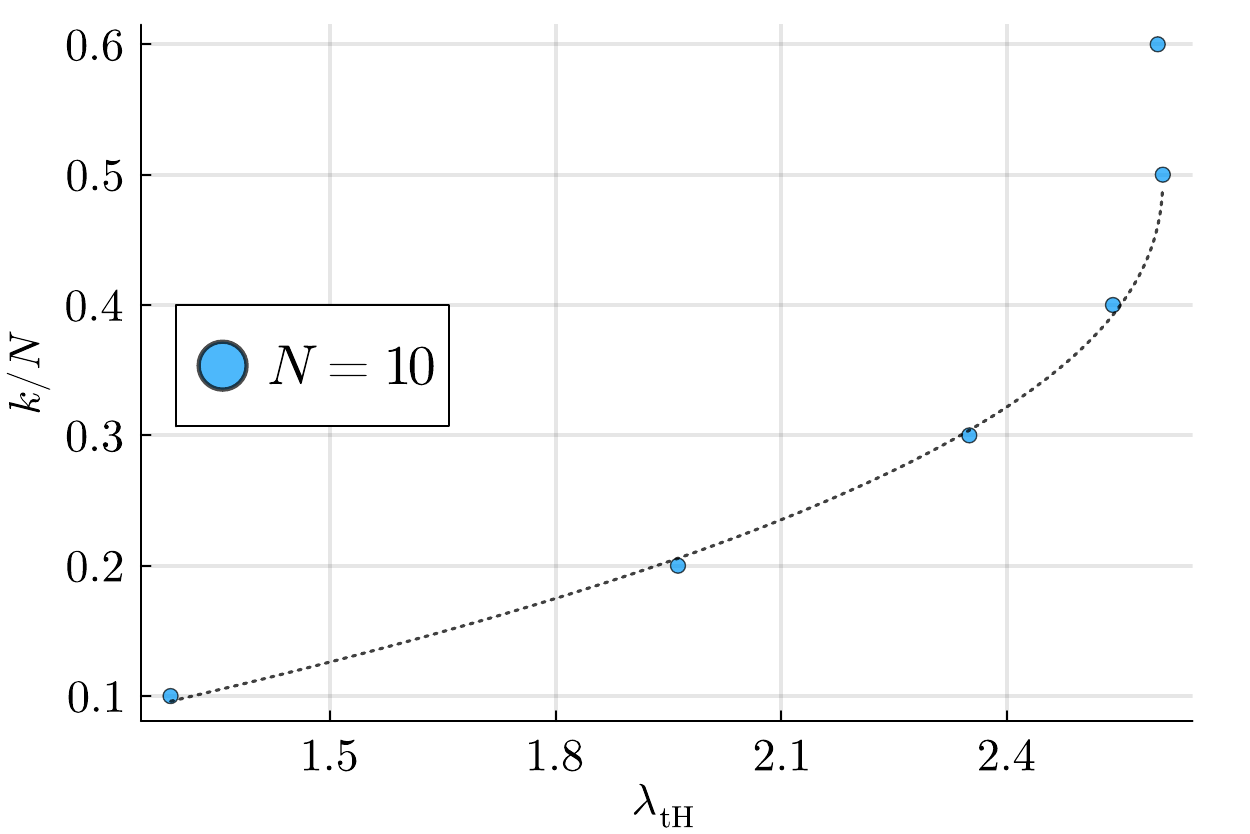}
    \caption{
    Phase diagrams in terms of the 't~Hooft coupling $\lambda_\mathrm{tH}$ and the ratio $k/N$. 
    The dotted lines represent the fitting results in the region $k\le k_\mathrm{ext}$ with the ansatz~\eqref{eq:fitting_ansatz_lower}.
    }
    \label{fig:lambda_vsk/N_lower}
\end{figure*}

\section{Fittings of phase boundaries}

This section presents the fitting results for phase boundaries at various integer values of $N$.
Figure~\ref{fig:lambdainv_vsk/N_upper} shows the phase diagrams in terms of the inverse 't~Hooft coupling $1/\lambda_\mathrm{tH}$ and the ratio $k/N$ for $N=2,\dots,7$, together with fitting curves obtained using the 4-parameter ansatz in Eq.~(9) of this Letter.
Since the fusion coefficients have been computed only for $|P_k| \lesssim 3000$, the number of data points available for the fitting is limited in the region $k\ge k_\mathrm{ext}(N) = k\big(\min_k\,1/g_\mathrm{c}^2(N,k)\big)$.
The corresponding fitting parameters for each $N$ are summarized in Tab.~II of this Letter.

We have also analyzed the behavior of the phase boundaries in the lower region $k\le k_\mathrm{ext}$.
In this case, we perform a fitting analysis for $N=6,\cdots,10$ in the $(\lambda_\mathrm{tH},k/N)$-plane by the following ansatz:
\begin{equation}
    \frac{k(\lambda_\mathrm{tH})}{N} 
    = 
    -\tilde A\qty(\lambda_{\max} - \lambda_\mathrm{tH})^{\tilde B} + \frac{k_\mathrm{ext}}{N},
    \label{eq:fitting_ansatz_lower}    
\end{equation}
where we have defined $\lambda_{\max} \coloneqq N g_\mathrm{c}^2(N,k_\mathrm{ext})$ and introduced two fitting parameters, $\tilde A$ and $\tilde B$.
Note that this ansatz cannot be transformed into the previous one for $k/N \gtrsim 0.5$ by a change of variables.
As shown in Fig.~\ref{fig:lambda_vsk/N_lower}, the points in the region $k/N \lesssim 0.5$ are well described by the fitting curves represented by the dotted lines.
The results of our best fits are summarized in Tab.~\ref{tab:fitting_result_lower}.

\begin{table}[th]
    \centering
    \caption{Fitting results of the phase boundaries in the region $k \le k_\mathrm{ext}$ in $(\lambda_\mathrm{tH},k/N)$-plane by Eq.~\eqref{eq:fitting_ansatz_lower}.}
    \begin{tabular}{|ccc|cc|}
        \hline
        $N$ &$k_\mathrm{ext}$ &$\lambda_{\max}$    &$\tilde A$   &$\tilde B$ 
        \\ \hline
        6   &3      &2.661       &0.3853164(1) &0.4010967(4)   \\
        7   &4      &2.637       &0.440(7)     &0.30(1)   \\
        8   &4      &2.627       &0.3667(4)    &0.44(1)   \\
        9   &5      &2.620       &0.4159(9)    &0.351(2)   \\
        10  &5      &2.608       &0.3574(4)    &0.44(2)   \\
        \hline
    \end{tabular}
    \label{tab:fitting_result_lower}
\end{table}

\bibliographystyle{utphys}
\bibliography{ref.bib}